\documentclass[a4paper,aps,superscriptaddress,floatfix,nofootinbib,twocolumn]{revtex4-1}

\usepackage{bbold}
\usepackage{bbm}
\usepackage[pdftex]{graphicx}
\usepackage{latexsym,amsmath,verbatim,amssymb,txfonts}
\usepackage{color}
\usepackage{rotating}
\usepackage{verbatim}
\usepackage{multirow}
\usepackage[english]{babel}
\usepackage{comment}
\usepackage{hyperref}

\usepackage{xcolor} 

\newcommand{\defeq}{\mathrel{\mathop:}=}
\newcommand{\SM}{~\cite{SMat}}
\newcommand{\av}[1]{\langle {#1} \rangle}

\begin{document}

\title{Classes of critical avalanche dynamics in complex networks}

\author{Filippo Radicchi}
\affiliation{Center for Complex Networks and Systems Research, Luddy School
  of Informatics, Computing, and Engineering, Indiana University, Bloomington,
  Indiana 47408, USA}
\email{filiradi@indiana.edu}

\author{Claudio Castellano}
\affiliation{Istituto dei Sistemi Complessi (ISC-CNR), Via dei Taurini 19, I-00185 Roma, Italy}

\author{Alessandro Flammini}
\affiliation{Center for Complex Networks and Systems Research, Luddy School
  of Informatics, Computing, and Engineering, Indiana University, Bloomington,
  Indiana 47408, USA}

\author{Miguel A. Mu\~noz} 
\affiliation{Departamento de Electromagnetismo y
  F{\'\i}sica de la Materia e Instituto Carlos I de F{\'\i}sica
  Te\'orica y Computacional. Facultad de Ciencias. Universidad de Granada.  E-18071,
  Granada, Spain}

\author{Daniele Notarmuzi}
\affiliation{Center for Complex Networks and Systems Research, Luddy School
  of Informatics, Computing, and Engineering, Indiana University, Bloomington,
  Indiana 47408, USA}

\begin{abstract}
  Dynamical processes exhibiting absorbing states are essential in the
  modeling of a large variety of situations from material science to
  epidemiology and social sciences. Such processes exhibit the
  possibility of avalanching behavior upon slow driving. Here, we
  study the distribution of sizes and durations of avalanches for
  well-known dynamical processes on complex networks. We find that all
  analyzed models display a similar critical behavior, characterized
  by the presence of two distinct regimes. At small scales, sizes and
  durations of avalanches exhibit distributions that are dependent on
  the network topology and the model dynamics. At asymptotically large
  scales instead --irrespective of the type of dynamics and of the
  topology of the underlying network-- sizes and durations of
  avalanches are characterized by power-law distributions with the
  exponents of the standard mean-field critical branching process.
\end{abstract}

\maketitle


\section{Introduction}

\begin{figure*}[!htb]
  \begin{center}
    \includegraphics[width=0.95\textwidth]{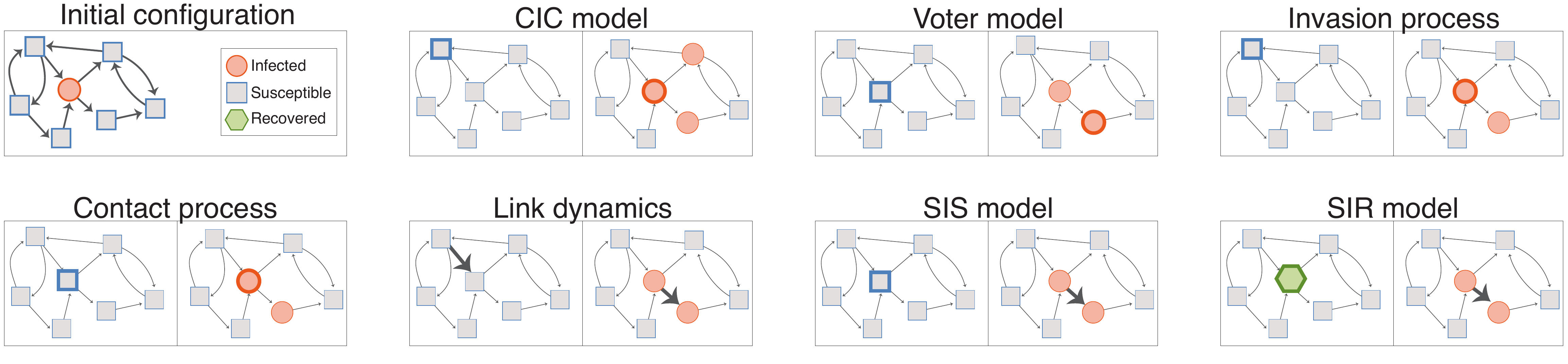}
  \end{center}
  \caption{Models for avalanche dynamics on networks. The figure
    serves as a schematic illustration to emphasize differences and
    similarities between the various dynamical models considered in
    this paper. The upper-left panel depicts an initial configuration,
    where a single node is in the ``active'' state, whereas all other
    nodes are inactive.  The rest of the panels display configurations
    reachable after one elementary reaction.  Depending on the model,
    elementary reactions are triggered by randomly selected nodes,
    edges, or both.  The elements triggering the elementary reactions
    are highlighted in the various panels. For clarity of the
    illustration, we report for each model only two of the possible
    configurations that can be reached after one elementary
    reaction. Specifically, configurations appearing in the left
    panels are reached after a recovery reaction, whereas
    configurations appearing in the right panels are obtained after a
    spreading reaction (in both cases the nodes or links triggering
    the reaction are highlighted with thick lines).  Detailed
    definitions of all dynamical models can be found in \SM.}
  \label{fig:1}
\end{figure*}


In this paper, we study seven stochastic models that are prototypical
to describe the diffusion of some sort of ``activity'' in
networks~\cite{CC-rmp, pastor2015epidemic}.  Specifically, our
analysis includes: the competition-induced-criticality model
(CIC)~\cite{Pinto, gleeson2014competition,notarmuzi2018analytical},
the voter model (VOT)~\cite{liggett1997stochastic,Krapivsky}, the
invasion process (IP)~\cite{castellano2005effect}, link dynamics
(LD)~\cite{sood2008voter}, the contact process
(CP)~\cite{durrett1984oriented}, the susceptible-infected-susceptible
model (SIS) and the susceptible-infected-recovered model
(SIR)~\cite{pastor2015epidemic} (see Figure \ref{fig:1}).  In all
these models, active nodes can pass the active status to their
inactive neighbors in the network, and can return to the inactive
status either spontaneously or by interacting with inactive
neighbors. The rules that govern these transitions are model
specific. All of the models are characterized by the existence of one
or more absorbing states, at which all dynamics ceases and the system
state remains frozen.  Typically, when parameters are set such that
the system is at the interface between the active and
inactive/absorbing phases, critical behavior --characterized by
power-law distributions of the sizes and durations of activity
avalanches-- emerges.

Many empirical studies of spreading phenomena in social, technological
and biological networks~\cite{onnela2010spontaneous,
  lerman2012social,MAM-rmp} reveal such a critical behavior. These
observations have triggered interest in understanding the origin of
criticality in specific dynamical models and its relationship with the
underlying network architecture.  The existing literature reveals that
the statistical properties of avalanches in some of the above models
may be dependent on the topology of the network on top of which the
dynamics proceeds~\cite{gleeson2014competition, NewmanSW,
  cohen2002percolation, Dorogovtsev08}, while some other authors
suggest that critical features are independent of network topology
\cite{larremore2012statistical}.  Here, we aim at reconsidering the
problem of critical scaling behavior on networks, for all the above
dynamical models within a  common and coherent perspective.


In many cases, the dependence of avalanche statistics on the topology
of the underlying network is theoretically explained by regarding the
avalanche as the result of a simple branching process
(BP)~\cite{Watson,Harris,Feller,Liggett}, where the network out-degree
distribution $P(k^{out})$ is identified with the distribution of
offspring number.  Such a mapping is exact as long as the evolution of
an avalanche does not substantially change the probability for an
active node to find inactive neighbors to infect, e.g., when the
substrate is a directed tree. However, the mapping usually fails for
arbitrary networks: after some transient time an active site may find
neighbors that are already active, so that the process is not merely
branching out, but interfering with itself.  Reasonably, this failure
is more dramatic in undirected networks as the front of an avalanche
can immediately move backwards and, as a consequence, break the
equivalence with a simple branching process.

According to the standard BP theory, sizes $S$ and durations $T$ of
avalanches at criticality --i.e., when on average there is one
offspring per active node-- are distributed according to power laws
\begin{equation}
  P(S) \sim S^{-\tau} {\cal{G_S}}(S/S_C) \; ~~\textrm{ and }~~ \; P(T) \sim
  T^{-\alpha} {\cal{G_T}}(T/T_C) \; 
\label{exponents}
\end{equation}
where $\tau$ and $\alpha$ are avalanche critical exponents, and
${\cal{G_S}}(S/S_C)$ and ${\cal{G_T}}(T/T_C)$ are cut-off (scaling)
functions, with the cut-off scales, $S_C$ and $T_C$, depending only on
system/network size right at the critical point~\cite{Leo,avalanches}.
Moreover, the average avalanche size scales with its duration as
$\langle S \rangle \sim T^\theta$, where the exponent $\theta$ obeys
the general scaling relationship
$\theta=(\alpha-1)/(\tau-1)$~\cite{Sethna,Colaiori1}.

The values of $\tau$ and $\alpha$ may depend on the offspring
distribution, i.e., the probability for an active node to activate a
given number of new nodes.  If the second moment of such a
distribution is finite, then
\begin{equation} 
\tau = 3/2, \; \; ~\alpha = 2 \;  \textrm{ and } \; ~ \theta = 2.
\label{eq:mf_exp}
\end{equation}
Eq.~(\ref{eq:mf_exp}) defines the so-called ``standard'' mean-field
(MF) or ``branching process'' exponents.  Thus, this type of
  scaling is expected to emerge for critical avalanches in the case in
  which the second moment of $P(k^{out})$ in the network is finite.

In fact, these values of $\tau$ and $\alpha$ are extremely universal
and robust; they emerge in many different types of propagation
processes such as directed percolation, CP, VOT, SIS, SIR and many
others, as long as the underlying pattern of connections is either a
high-dimensional lattice or a sufficiently homogeneous
network~\cite{survival1,avalanches,survival2,Redner,Henkel}~\footnote{A
  particularly simple proof of the emergence of the standard exponents
  when the underlying tree is homogeneous with $k^{out} = 2$ can be
  found in Ref.~\cite{diSanto2017simple}.  A more systematic
  derivation --for different types of underlying regular or random
  tree topologies-- can be obtained within the generating-function
  formalism~\cite{Redner,Plischke,SOBP}; for instance, already back in
  1949, Otter computed the solution when $P(k^{out})$ is a Poisson
  distribution~\cite{Otter}.}. This super-universality can be
rationalized using a Langevin equation for the density $\rho$ of
active sites
\begin{equation}
	\frac{d \rho(t)}{dt} = F \, \sqrt{ \rho(t) } \, \xi(t) 
	\label{Langevin}
\end{equation}
where $F$ is a noise-amplitude constant and $\xi(t)$ a zero-mean
Gaussian white noise, which is shared, as an effective mean-field
description, by all the above mentioned
models~\cite{diSanto2017simple}. Observe that the square-root term in Eq.~(\ref{Langevin})
accounts for ``demographic'' fluctuations and is a direct consequence
of the central limit theorem \cite{Henkel,Marro}.

  On the other hand, if the second moment of the offspring
  distribution --or, equivalently $P(k^{out})$-- diverges, then the
  critical exponents of the associated branching process differ from the standard
  MF ones of Eq.~(\ref{eq:mf_exp}). In particular, for
  $P(k^{out}) \sim [k^{out}]^{-\gamma}$ with $2<\gamma <3$, one
  obtains $\gamma$-dependent exponents \cite{Goh,Sornette,
    gleeson2014competition},
\begin{equation}
\tau =   \frac{\gamma}{\gamma-1} \;, \; \alpha =  \frac{\gamma-1}{\gamma-2} 
\;  ~\textrm{ and }~\theta = \frac{\gamma-1}{\gamma-2}.
\label{eq:anomalous_exp}
\end{equation}
Observe that these ``anomalous'' branching process (ABP) exponents
converge to those of Eq.~(\ref{eq:mf_exp}) in the limit
$\gamma \rightarrow 3$, i.e., when the second moment of $P(k^{out})$
becomes finite (with the caveat of logarithmic corrections to scaling
at the marginal value $\gamma=3$~\cite{Goh})~\footnote{Actually, the
  above-mentioned effective Langevin-equation approach breaks down for
  $\gamma<3$, as it includes a standard Gaussian noise, stemming
  ultimately from the central limit theorem for the addition of
  stochastic variables with finite variance. In the case $\gamma<3$,
  the variance of the variables to be added is not finite, the
  Gaussian noise needs to be replaced by a L{\'e}vy-stable
  distribution, leading to a different type of effective description
  and to anomalous behavior [i.e., Eq.(\ref{eq:anomalous_exp})]
  ~\cite{Sornette-book}.}.

Real-world networks often exhibit power-law degree distributions
$P(k) \sim k^{-\gamma}$ with
$2<\gamma <3$~\cite{barabasi1999emergence, voitalov2018scale}, and
studies concerning spreading processes often assume underlying
scale-free network topologies~\cite{CC-rmp,pastor2015epidemic}.  A
naive extension of the standard BP mapping to networks with diverging
second moment of the degree distribution suggests that one should
generically observe anomalous exponents.  Does anomalous scaling hold
for avalanches on real networks?

The current literature on the existence of anomalous avalanche scaling
in scale-free networks reports conclusions that are often contradictory
or difficult to reconcile with each other. For example, according to
Larremore {\it et al.}  critical avalanches on networks are always
characterized by the standard MF exponents of Eq.~(\ref{eq:mf_exp})
irrespective of the underlying network topology
~\cite{larremore2012statistical}.  However, numerical evidence in
support of such a claim is presented only for power-law networks with
degree exponent $\gamma > 3$.  Furthermore, this is in apparent
contradiction with what reported for the CIC model on directed
scale-free networks. In particular, Gleeson {\it et al.} employ a map
onto an anomalous branching process to argue that one should expect
anomalous scaling for $2< \gamma < 3$ and standard MF critical
exponents for $\gamma > 3$~\cite{gleeson2014competition}, but offered
limited computational evidence in support of such a claim.

Also for the broadly studied SIR model the current state of
understanding is not entirely clear.  First, the model is never
studied directly; rather, claims follow from SIR equivalence with bond
percolation~\cite{grassberger1983critical}, according to which the
distribution of SIR avalanche sizes can be deduced from the
percolation cluster-size distribution. Theoretical claims on bond
percolation in scale-free networks mostly regard undirected
networks~\cite{NewmanSW, cohen2002percolation, Dorogovtsev08}.  This
is a very difficult setting to consider given that the percolation
threshold vanishes~\cite{cohen2000resilience}.  A large-scale
numerical study of the percolation cluster size distribution in
scale-free graphs is the one of Ref.~\cite{radicchi2015breaking},
where critical exponents seem compatible with the standard ones of
Eq.~(\ref{eq:mf_exp}) for any $\gamma > 2$.

The goal of this paper is to provide a coherent picture for avalanche
statistics in critical processes taking place on networks.  The study
consists in extensive numerical simulations, combined with analytical
arguments, of the various avalanche models on a variety of networks,
both synthetic and real.

\section{Models: networks and dynamics}
The models studied here are described in detail in \SM.
Figure~\ref{fig:1} illustrates schematically the mechanisms at the
basis of the various models under consideration.  As a substrate for
the dynamics of activity, we assume in all cases a network composed of
$N$ nodes. The topology of the network is fully specified by its 
adjacency matrix $A$, whose generic element $A_{ij} = 1$ if an edge
from node $i$ to node $j$ exists and $A_{ij} = 0$ otherwise. We assume
that no selfloops are present in the network, i.e., $A_{ii} =0$ for
all $i$.  In the most general case, we consider directed networks,
where $A_{ij} \neq A_{ji}$.
For simplicity, we further assume that the network is composed of only
one strongly connected component, so that at least one directed path
between any pair of nodes exists. 
  
The state of the system at time $t$ is denoted by the vector
$\vec{\sigma}(t) = [\sigma_1(t), \sigma_2(t), \ldots, \sigma_N(t)]^T$,
where $\sigma_i(t)$ is a discrete-valued variable representing the
state of node $i$ at time $t$. In all models except for the SIR,
$\sigma_i(t)$ can assume two values: $\sigma_i(t) = 1,0$ indicating
that the node is active, inactive respectively. In the SIR one can
also have that $\sigma_i(t) \neq 0, 1$ meaning that node $i$ is
recovered and does not participate any longer in the dynamics.

All models are stochastic Markov processes where the elementary
reactions that lead to changes in system configurations are triggered
by the random selection of network elements, either nodes,
edges or both. Propensities of the various reactions may depend on exogenous
parameters whose values can be tuned to bring the system into
different dynamical regimes.
All models are characterized by
  an asynchronous updating scheme, meaning that an elementary dynamical step
  leads to the change of the state of at most all neighbors of a
  single node.
The state
$\vec{\sigma} = \vec{0} = (0, 0, \ldots, 0)^T$ is an absorbing
configuration for all models.  Additional absorbing configurations are
present in some models. For example, in the SIR model, all
configurations with no infected nodes, but arbitrary number of
recovered nodes are absorbing; in some other models, such as the CIC,
the configuration $\vec{\sigma} = \vec{1} = (1, 1, \ldots, 1)^T$ is
also absorbing at the critical point.

We are interested in the critical regimes of the considered dynamical
models.  The criterion to achieve criticality is model specific.  VOT,
IP and LD have no free parameters and are intrinsically critical. CIC
critical point is achieved by setting model parameters to
network-independent values. For CP a known network-independent value
is a good approximation of it. The critical point of SIS is
approximated by considering the inverse of the largest eigenvalue of
the adjacency matrix of the
graph~\cite{goltsev2012localization,larremore2012statistical}.  For
SIR, the critical regime is approximated relying on the value of the
largest eigenvalue of the non-backtracking matrix of the
graph~\cite{karrer2014percolation, radicchi2015predicting} (see
\SM~for further details).

The elementary rules at the basis of the various
  dynamical models are described in detail in the SM.
  Here, we briefly illustrate such rules for the sake of clarity. We
  remind that the underlying network is potentially directed, so that $A_{ij} = 1$
  indicates the existence of the connection $i \to j$. In the critical CIC, a
  randomly chosen node $i$ shares its state with all its
  neighbors, i.e., $\sigma_j(t+dt) = \sigma_i(t)$ for all $j$ such that $A_{ij} =
  1$, where $dt$ indicates the amount of time needed for
  the elementary reaction to occur. In the VOT model, a randomly chosen node $i$ inherits the state
  of one randomly chosen neighbor $j$, $\sigma_i(t+dt) = \sigma_j(t)$ with
 $A_{ji} =1$. In the IP, a randomly chosen node $i$ copies its own state on a randomly chosen
 neighbor $j$, $\sigma_j(t+dt) = \sigma_i(t)$ with $A_{ij} =1$.
 In the LD, a random edge $i \to j$ is first selected,  then the state of 
  node $j$ becomes identical to the one of node $i$, $\sigma_j(t+dt) =
  \sigma_i(t)$.  In the CP, two possible events may happen: (i) a randomly chosen
  active node $i$ may recover, i.e., $\sigma_i(t) = 1 \to \sigma_i(t+dt) = 0$; (ii) a
  randomly chosen node $i$ spreads its activity on a randomly chosen
  neighbor $j$, i.e., $\sigma_i(t) = 1$ and $A_{ij} =1$
  causes $\sigma_{j}(t+dt) = \sigma_i(t)$.
  Also in the SIS  two possible events may occur: (i)  a randomly chosen
  active node $i$ may recover, i.e., $\sigma_i(t) = 1 \to \sigma_i(t+dt) = 0$; (ii) a
  random link $i \to j$ is chosen so that $\sigma_i(t) = 1$ and
  $\sigma_j(t) = 0$, then $\sigma_j(t) = 0 \to \sigma_{j}(t+dt) = 1$.
  Finally, the rules of the SIR are almost the same as those of SIS. The only
  difference is that the recovery event (i) leads to the change
  $\sigma_i(t) = 1 \to \sigma_i(t+dt) \neq 0, 1$, and nodes in the recovered
  state do not longer participate in the dynamics.

It is possible to classify the
  various models in three main classes of dynamical behavior.  The
  first class is formed by the CP
  and IP, for which the expected number of spreading events in which the generic
  node $i$ influences its neighbors is a constant independent of the
  out-degree $k^{out}_i$. 
They differ from the CIC, VOT, LD, SIS and SIR for which the expected number of spreading
  events in which node $i$ influences its neighbors is directly
  proportional to the out-degree $k^{out}_i$. 
As stressed above, the CIC, VOT and LD are tuned to criticality in a way that is independent of
  the underlying network topology, thus they constitute a separate
  class from the one of the SIS and SIR models, whose critical
  regime is determined by the network topology.

We consider avalanches initiated by a single randomly-chosen node,
$j$, so that the initial configuration is $\sigma_i(0) = 0$ for all
$i \neq j$, and $\sigma_j(0) = 1$.  We follow the dynamics of each
avalanche until the system reaches an absorbing configuration.  We
define the duration $T$ as the number of time steps needed to reach an
absorbing configuration. We also define the size $S$ as the number of
elementary spreading events occurred during $T$. An elementary
spreading event is the occurrence of an active node passing activation
to other, not necessarily inactive, nodes.

All nodes may participate multiple times in an avalanche (i.e., they
can be ``re-activated'') so that the network size $N$ is not an upper
bound for $S$; the only exception to this rule is the SIR model where
nodes can be activated only once.  We focus on finite avalanches only,
i.e., those that end up in absorbing configuration
$\vec{\sigma} = \vec{0}$. In the CIC model, for instance, we exclude
avalanches that end in $\vec{\sigma} = \vec{1}$, as they can be viewed
as infinite avalanches. For the SIR model, we consider instead all
avalanches.

\begin{figure*}[!htb]
  \begin{center}
    \includegraphics[width=0.95\textwidth]{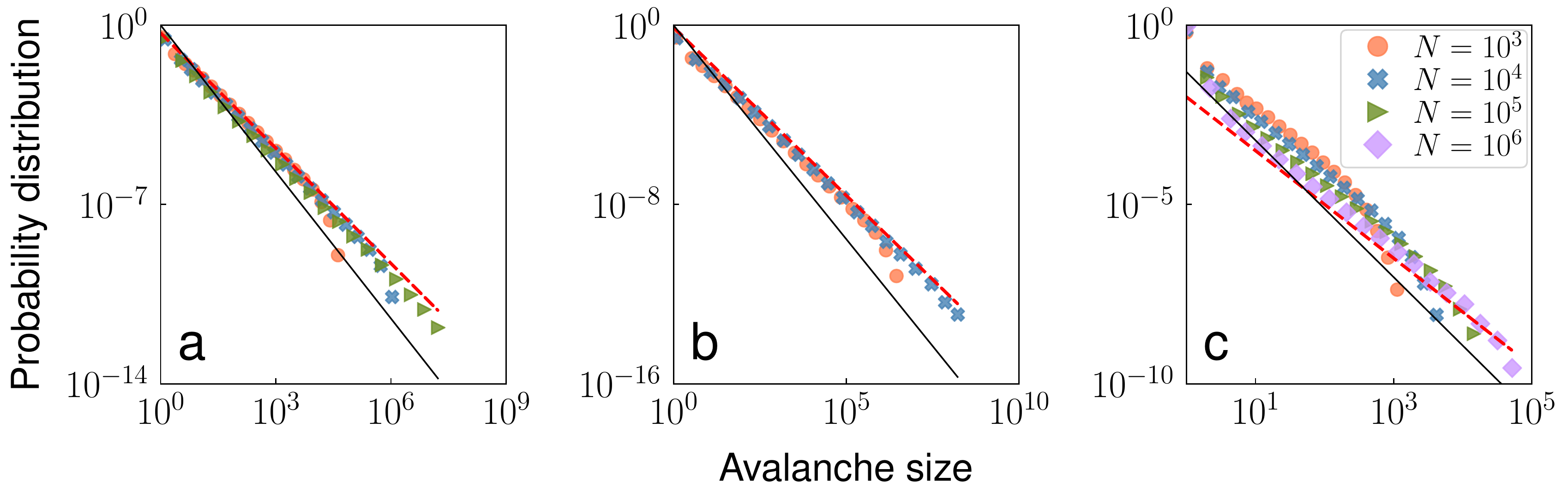}
  \end{center}
  \caption{Avalanche size in synthetic undirected scale-free networks.
    The degree exponent is $\gamma = 2.1$.  For clarity, we report
    results only for three dynamical models: (a) CIC, (b) IP, and (c) SIS.
    Results for all other models are in \SM.~ For each model and
    network, we measure, by means of numerical simulations, the
    probability distribution $P(S)$ of the total number of spreading
    events $S$ per avalanche.  The dashed red line corresponds to MF
    exponents, i.e., Eq.~(\ref{eq:mf_exp}); the full black line
    indicates anomalous BP scaling i.e.,
    Eq.~(\ref{eq:anomalous_exp}). }
  \label{fig:2}
\end{figure*}

We study, by means of extensive computational simulations, avalanche
statistics for all the above-mentioned prototypical dynamical models
on top of scale-free networks generated by one of two possible standard
generative models, both of which produce uncorrelated random graphs
with power-law degree distributions.

First, we consider undirected scale-free networks obtained via the
uncorrelated configuration model~\cite{catanzaro2005generation} with
degree distribution $P(k) \sim k^{-\gamma}$ with support
$ [4, \sqrt{N}]$.  In our numerical analyses, we set
$\gamma = 2.1$ and vary the network size $N$.  The choice
$\gamma = 2.1$ is expedient because it corresponds to a large gap
between standard MF and ABP exponents, easing computational
  discrimination of scaling regimes. Results for $\gamma = 2.5$ are
reported in \SM.  We generated a single graph instance of the
model for every $N$, and used these graphs in all our
analyses. We tested that choosing a particular instance of the graph model does not
affect the statistics of avalanches. 
\footnote{Please note that each network size corresponds indeed to a different
instance of the network model. Observe that no significant variation
among the various network instances is visible, if not due to
finite-size effects.}
For
every network and model, we simulated $10^6$ avalanches seeded at a
randomly chosen single node and measured the corresponding avalanche
size and duration distributions, i.e., $P(S)$ and $P(T)$,
respectively.

The first major result of our analyses (see Fig.~\ref{fig:2}) is that
--when deployed on undirected scale-free networks-- all activation
models considered are characterized asymptotically (for large $S$ and
large $T$) by standard MF exponents.  This happens regardless of the
fact that we have set $2 < \gamma < 3$, i.e., for networks for
which a strict analogy with branching process would suggest anomalous
exponents.  An exhaustive report of the results of our analysis is
shown in \SM.

\begin{figure*}[!htb]
  \begin{center}
    \includegraphics[width=0.95\textwidth]{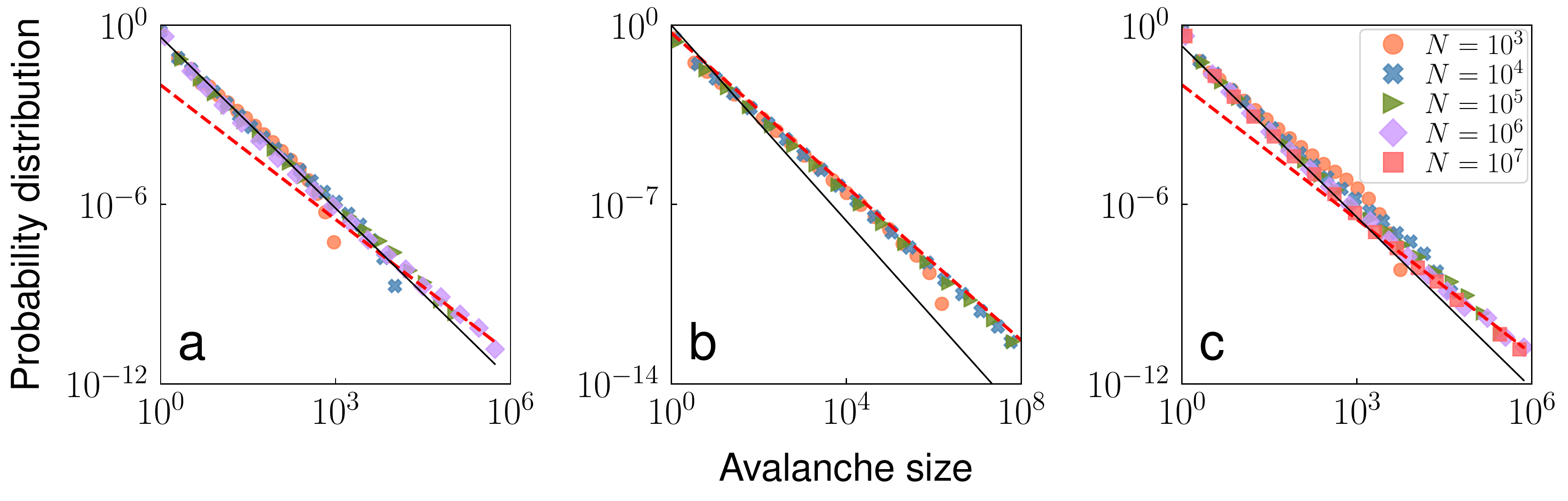}
  \end{center}
  \caption{Avalanche size in synthetic directed networks for three of
    the considered models: (a) CIC, (b) IP, and (c) SIS. The description
    of the figure panels is as in Fig.~\ref{fig:2}. The
    networks analyzed here are instances of the directed configuration
    model with out-degree exponent $\gamma = 2.1$ and maximum out-degree
    $k_{max}^{out} = N-1$. }
  \label{fig:3}
\end{figure*}

Second, we analyze directed scale-free networks constructed according
to the model of Ref.~\cite{gleeson2014competition}. This is a simple
extension of the configuration model to generate directed networks,
where node out-degrees $k^{out}$ are drawn from the distribution
$P(k^{out}) \sim [k^{out}]^{-\gamma}$ for
$k^{out} \in [4, k^{out}_{max}]$ and $P(k^{out}) = 0$, otherwise
\footnote{In the generation of a graph instance, each node $i$ is
  connected to $k^{out}_i$ other nodes, chosen at random in the
  network, so that in-degrees obey a Poissonian distribution with
  average value equal to the average out-degree.}  (see \SM).  We
consider $\gamma = 2.1$ as above (results for $\gamma = 2.5$ are also
reported in \SM) and set either $k^{out}_{max}=\sqrt{N}$ or
$k^{out}_{max}= N-1$.

When $k^{out}_{max} = \sqrt{N}$ our simulations show that all the
activation processes are again in the MF universality class (see \SM)
for all values of $\gamma$.

A more complicated scenario emerges when $k^{out}_{max} = N-1$. In
Fig.~\ref{fig:3}, we show the results only for few selected
models. Results for all other models are reported in \SM.  For IP and
CP, we still observe clear MF scaling, in all networks.  In all other
models, the distribution of the avalanche sizes displays a crossover
from anomalous (for small sizes/durations) to MF exponents (for large
sizes/durations). The crossover point increases with the network size,
suggesting that only anomalous exponents should be present in the
limit of asymptotically large networks~\footnote{Let us stress that
  the observed scaling of $P(T)$ and power-law relation between $S$
  and $T$ provide much less clear evidence of anomalous BP behavior
  even for the case $k^{out}_{max} = N-1$ (see \SM). This issue is due
  to the finite size of the networks, and it is visible also in
  numerical results concerning pure BP with finite-size constraints
  \SM. Clearer observations of anomalous critical exponents for
  $P(T)$ and the power-law relation between $\langle S \rangle$ and
  $T$ can be obtained for $\gamma = 2.5$; such a choice of the
  $\gamma$ value leads however to much less noticeable differences
  between anomalous and standard exponents for the distribution $P(S)$
  \SM.}.

In summary, our results provide strong support for MF exponents in all
situations where a priori we expect standard BP behavior (i.e., finite
second moment of the out-degree distribution). Settings for which one
could predict \emph{a priori} anomalous BP scaling (i.e. for
scale-free networks with $2<\gamma<3$) generate results that are much
less cleancut.  Anomalous exponents can at most be observed only in
the regime of small avalanches for the distribution of the avalanche
size. Strong deviations from the predicted anomalous power-law
scalings are observed otherwise.  
An important role for the
  observation of anomalous exponents is played by the upper bound of
  the out-degree distribution.   The magnitude of the largest degree, and
  in general, the frequency of high degree nodes determines the
  quantity of  superspreaders and it is therefore a
  crucial quantity to consider in diffusion models. The upper bound determines how fast the
  second moment of the distribution diverges as the network size is
  increased. On unweighted networks, maximal divergence is obtained
  for $k_{max} = N-1$, corresponding to the setting where anomalous
  scaling can be best appreciated. Slow divergence, as for the case
  $k^{out}_{max} = \sqrt{N}$, makes it difficult to observe anomalous
  behavior, at least for the size of the networks that we are
  considering in this work.

\section{Analytical approach}

Two alternative types of approaches are frequently used to study
avalanches in networks. The first one is, as discussed above, a
branching process approximation for cascades of a small size at the
beginning of the process. The second one is an approximation by a
dynamical system once the avalanche spreads to a significant fraction
of the underlying network~\cite{Rastegar}.  The mathematical approach
we develop in what follows belongs to this second group.

Depending on their specific features, all dynamical models under
consideration can be grouped into three classes described by different
types of mathematical equations.  These groups are: (i) IP and CP;
(ii) CIC, VOT, and LD; (iii) SIS and SIR.

The first group is trivially described by dynamical processes that are
insensitive to the out-degree sequence of the underlying network. In
other words, anomalous propagation events in which a single active
node propagates activity to an arbitrarily large number of nearest
neighbors are simply not allowed by the dynamics.  Henceforth,
anomalous type of scaling is not expected to appear, even at the level
of a naive mapping onto a branching process.  Thus, IP and CP
avalanches are expected to be always characterized by standard MF
exponents (see \SM), in perfect agreement with our computational
results.

Models in the other two classes have instead a much less trivial
behaviour.  We consider the CIC and SIS models as representative of
each of these two classes and derive a mathematical approach for
each. The full development of the theory (and extension to the other
dynamical models) is presented in \SM; here, we sketch the main
results and the main insights derived from them.

Our analytical approach is based on two successive approximations.
The first one is the so-called individual-based mean-field
approximation (IBMFA) (see ~\citet{pastor2015epidemic} for a
review). This analysis starts by describing the evolution of the
average value of the state of an individual node in the network, where
the average is taken over many realizations of the dynamical process.
The approximation consists in neglecting dynamical correlations among
variables, so that every node feels only the influence of the average
behavior of each of its neighbors. We use the IBMFA for determining
how and when the system reaches its long-term dynamical regime.  The
second approximation consists in deriving a Langevin equation for the
overall network activity, written as the sum of the activity variables
of all nodes ~\cite{boguna2009langevin}.  From this approximation, it
is possible to derive the statistics of long-term avalanches based on
the equivalence between the resulting Langevin equations and
Eq.~(\ref{Langevin}), derived in Ref.~\cite{diSanto2017simple} to
describe the standard branching process and related processes.

Let us first present the derivation of the main results for CIC
critical dynamics, in which the only possible change in the state of
a node consists in copying the state of a nearest neighbor. In the IBMFA,
we focus our attention on the deterministic node variable
$s_i(t) \defeq \langle \sigma_i(t) \rangle$, defined as the value of
the stochastic variable $\sigma_i(t)$, averaged over the realizations
of the dynamical process at time $t$.  As we explicitly derive in the
\SM, critical CIC dynamics is described by the IBMFA equation
\begin{equation}
	\frac{d \vec{s} (t)}{dt} = 
          - L^T \, \vec{s}(t) 
\; .
	\label{eq:IBMFA_CIC}
\end{equation}
Here, $\vec{s(t)} = [s_1(t), s_2(t), \ldots, s_N(t)]^T$ is the
vector describing the average state of the nodes of the network at
time $t$.  $L = K^{in} - A$ is the (directed) graph Laplacian of the
network, with $K^{in}$ the diagonal matrix whose non-null elements are
equal to the in-degree of the nodes, and $A$ is the graph adjacency
matrix \cite{caughman2006kernels}.  In essence, under the IBMFA, the
critical CIC coincides with a purely diffusive process. The properties
of the solutions of Eq.~(\ref{eq:IBMFA_CIC}) for arbitrary graphs are
described in Refs.~\cite{masuda2017random, veerman2018diffusion}; we
briefly summarize them here. If the underlying network is composed of
a single strongly connected component, then, the long-term behavior is
such that $\lim_{t \to \infty} s_i(t) = s^*$, for all nodes
$i$. Because the asymptotic limit of the individual variables $s_i$
does not depend on $i$, $s^*$ coincides with the asymptotic value of
density, $r^*$. The latter is given by the norm of the vector $s(t)$
at large times, i.e.,
$r^* = \lVert [ (\vec{v}_1^{(l)})^T \cdot \vec{s}(t=0) ] \cdot
\vec{v}_1^{(r)} \rVert $, where $\vec{v}_1^{(l)}$ and
$\vec{v}_1^{(r)}$ are the left and right eigenvectors of $L^T$,
respectively. These eigenvectors correspond to the smallest 
eigenvalue $\nu_1 =0$ of the graph Laplacian.  The long-term regime is reached
exponentially fast.  However, depending on the type of network, we
have different diffusion behaviors. For undirected networks,
$\vec{v}_1^{(l)} = \vec{v}_1^{(r)} = \vec{1} / \sqrt{N}$; further, the
density of active nodes $r(t) \defeq 1/N \, \sum_i s_i(t) $ is such
that $r(t) = r(0) = 1/N$ for all $t$. The typical time scale is
$t^* = 1/ \nu_2$, with $\nu_2$ the second smallest eigenvalue of $L$
(see \SM).

On the other hand, if the network is directed,
  $\vec{v}_1^{(r)} = \vec{1} / \sqrt{N}$, but
  $\vec{v}_1^{(l)} \neq \vec{v}_1^{(r)}$. This means that, also in
  this case the vector $\vec{s}$ has identical components for
  $t \to \infty$. However, $r(t)$ may increase or
decrease depending on the initial condition, so that the steady-state
value $r^*$ of the density is sensitive to the initial choice of the
seed node~\footnote{For instance, in \SM~ we show that for the
  directed configuration model, the $i$-th component of such a vector
  is proportional to the out-degree of node $i$, i.e.,
  $v_{1,i}^{(l)} \sim k^{out}_i$.}. Further, in this case, we cannot
longer apply the spectral theorem to the corresponding non-symmetric
matrix so that the relaxation to the steady-state cannot
be easily written in terms of the Laplacian eigenvalues.


To determine the statistical properties of avalanches with duration
$T \gg t^*$, i.e. asymptotically, we now go back to the stochastic
description of the full dynamical system. We take advantage of the
previous finding obtained under the IBMFA, and assume that
$\rho(t) \defeq 1/N \, \sum_i \sigma_i(t)$ is a quantity that
fluctuates around its average value $\av{\rho}=r^*$. Essentially, we
make the hypothesis that the system has reached a stationary state
where the number of active nodes is constant on average, but still
subjected to demographic fluctuations.  In analogy with
Ref.~\cite{boguna2009langevin}, we refer to this assumption as the
adiabatic approximation.  Thus, the dynamics of long-term avalanches in CIC
critical dynamics turns out to obey the following Langevin
equation (see \SM)
\begin{equation}
\frac{d \rho(t)}{dt} = \sqrt{2 \langle k^{in} \rangle / N} \, \sqrt{\rho(t) [1-
  \rho(t)]} \, \xi(t)\; ,
\label{eq:langevin_CIC}
\end{equation}
where $\xi(t)$ is a zero-mean Gaussian white noise, and
$\langle k^{in} \rangle$ is the average in-degree of the network.  The
dependence on $\rho$ of the diffusion coefficient imposes the absence
of fluctuations for both $\rho=0$ and $\rho=1$, corresponding to the
two existing absorbing states.  Except
for higher-order terms, Eq.~(\ref{eq:langevin_CIC}) has the generic
form of the representative Langevin equation, Eq.(\ref{Langevin}) for
avalanches in the class of standard MF branching processes
\cite{diSanto2017simple}.  This implies that long-term avalanches in
critical CIC dynamics obey power-law distributions with MF critical
exponents, i.e., Eqs.~(\ref{eq:mf_exp}).

We now briefly illustrate the analytical approach for SIS critical
dynamics. We basically repeat the same steps described above for
critical CIC dynamics.  The IBMFA equation reads as
\begin{equation}
	\frac{d \vec{s} (t)}{dt} = 
 (A^T -
 I)\, \vec{s} (t)  
\; ,
	\label{eq:INMFA_SIS}
\end{equation}
where
$I$ is the identity matrix~\cite{goltsev2012localization}.  The
solution of the IBMFA equation is a vector whose components are
proportional to those of the principal right eigenvector
$\vec{w}_N^{(r)}$ of the matrix $A^T$~\cite{goltsev2012localization};
convergence to the asymptotic solution is exponentially fast.  The
asymptotic value of density of active nodes is given by
$ r^* = \lVert (\vec{w}_N^{(l)} )^T \cdot \vec{s}(t=0) ] \,
\vec{w}_N^{(r)} \lVert$, with $\vec{w}_N^{(l)}$ and $\vec{w}_N^{(r)}$
principal left and right eigenvector of the matrix $A^T$
respectively~\footnote{In uncorrelated random network models, the
  components of the vector $\vec{w}_N^{(l)}$ are proportional to the
  node out-degrees, i.e., $w_{1,i}^{(l)} \sim k^{out}_i$~(see
  \SM). For undirected configuration models, the previous statement is
  valid only when the degree exponent $\gamma < 5/2$.}.  If the
network is undirected, the time scale of the exponential relaxation to
the steady-state density is given by
$t^* = \omega_N / (\omega_N - \omega_{N-1})$, with $\omega_N$ largest
eigenvalue of $A$, and $\omega_{N-1}$ second largest eigenvalue of $A$
(see \SM).  If the network is directed, $t^*$ is not directly
quantifiable in terms of the eigenvalues of the matrix $A$.

For $t \gg t^*$, the system has reached its long-term dynamical
regime. The statistics of long avalanches is described by the Langevin
equation
\begin{equation}
\frac{d \rho(t)}{dt} = \sqrt{2 \langle w_N^{(r)} \rangle / N } \, \sqrt{\rho(t)} \, \xi(t)\; .
\label{eq:langevin_SIS}
\end{equation}
where 
$\langle w_N^{(r)} \rangle$ is the average value of the components of the
principal right eigenvector of the matrix $A^T$ (see \SM). 
Eq.~(\ref{eq:langevin_SIS}) has the same 
form as those considered by di Santo {\it et al.}~\cite{diSanto2017simple}, 
valid for avalanche models that are equivalent to standard BP processes.
This tells us that long-term avalanches in critical SIS dynamics
obey power-law distributions with MF 
critical exponents, i.e., Eqs.~(\ref{eq:mf_exp}).

\begin{figure*}[!htb]
  \begin{center}
    \includegraphics[width=0.95\textwidth]{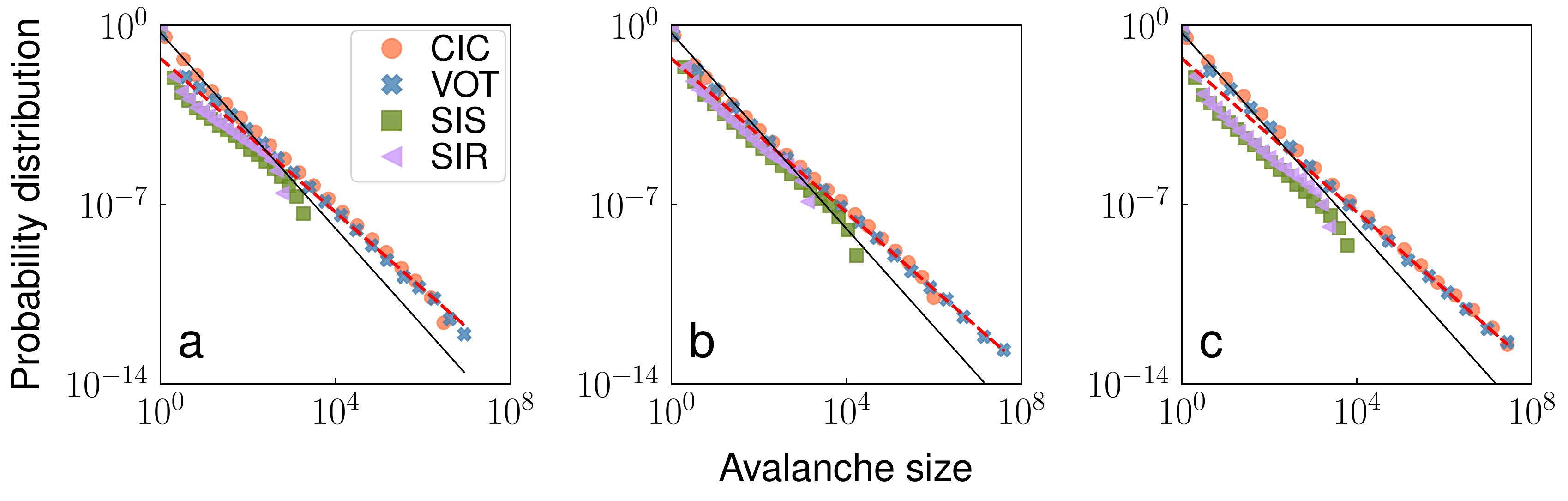}
  \end{center}
  \caption{Avalanche size in real-world networks.  We consider the
    following networks: (a) undirected graph representing a snapshot
    of the Internet at the Autonomous system
    level~\cite{leskovec2005graphs}; (b) directed Twitter network of
    the Spanish 15M movement~\cite{borge2011structural}; (c) directed
    graph representing a portion of the Youtube social
    network~\cite{mislove-2007-socialnetworks}.  Different symbols and
    colors refer to different avalanche dynamical models. The red
    dashed line represent standard BP critical exponents, while the
    full black line indicates the power-law decay expected for
    anomalous BP. Note that the out-degree distributions of these
    networks are all well modeled by power laws with decay exponent
    $\gamma = 2.1$ (see \SM). }
  \label{fig:4}
\end{figure*}

In summary, the above analytical approach tells us that sufficiently
long (large) avalanches in critical CIC and SIS dynamics should follow
a standard MF scaling. This conclusion, in principle is true for any
network.  However, an avalanche is sufficiently long to obey standard
BP statistics only if its duration is much longer than the typical
time scale that can be deduced from the IBMFA of the process happening
on the network.  The magnitude of such time scale depends exclusively
on the topology of the network, by means of either the Laplacian or
the adjacency matrix of the graph.  Undirected networks with
sufficiently short average distance, for instance, have a relatively
small diffusion time scale. There are, however, network topologies
where diffusion may be particularly slow to reach its stationary
state. Examples are low-dimensional lattices, and networks with long
loops. In these cases, the vast majority of observed avalanches may
never be long enough as to be describable by the long-term statistics.
We do not have analytical arguments to determine the statistical
properties of avalanches in the short-term dynamical regime, but, in
principle, one expects that the effective mapping into an ABP should
work (for scale-free networks with $2<\gamma<3$).  Our numerical
results seem to indicate that anomalous BP scaling is possible as long
as the underlying networks are directed and have power-law out-degree
distributions. The cutoff of the out-degree distribution seems also to
play an important role for the possible observation of anomalous
scaling, at least for the network sizes that we were able to consider
in our analysis.

\section{Real networks}
All considerations, numerical and theoretical, made for synthetic
graphs are valid also for real-world networks.  In Fig.~\ref{fig:4},
we summarize the results of numerical simulations performed on three
large-scale networks. Additional results are provided in \SM.  A
pre-asymptotic regime with anomalous scaling for sufficiently small
avalanches is seen for example in the Youtube direct social network
(Fig.~\ref{fig:4}c).  The distributions for large avalanches are
instead very well described by MF critical exponents in all cases.
The crossover
point may be interpreted as the typical scale that distinguishes local
from global avalanches, and it could be employed as a quantitative
criterion to tell whether an avalanche is ``viral'' or not.

\section{Conclusions and discussion}
In conclusion, we found that any minimal deviation from the
assumptions underlying the mapping into an anomalous branching process
brings the system back to the realm of standard MF and its associated
super-universal exponents, so that anomalous exponents are exceedingly
difficult to observe.  
Our results suggest
this statement
to be true for
seven well-known avalanche dynamical models, but we believe that it
can be extended to many other spreading processes taking place on
networks.
 Our results are valid for avalanche dynamical models with
   asynchronous updating rules. We do not exclude that models with synchronous
   updates may exhibit different statistical properties, with anomalous
   exponents emerging even in settings less peculiar than those
   identified here for asynchronous dynamical models~\cite{Zierenberg}.

Why is numerical evidence of anomalous scaling so weak, even in the
cases when intuition suggests that the dynamical avalanche model could
be well mapped to an anomalous branching process?  Clearly, if the
process is occurring on a directed tree with power-law out-degree
distribution, then anomalous scaling occurs. However, avalanche models
in more complex networks do not necessarily satisfy such strict
conditions.  There are many possible ways in which the assumptions of
the mapping to an anomalous branching process can be violated.

First, both in directed and undirected networks avalanches do not
necessarily proceed in a fully feedforward way; already active nodes
can be found by a branch of an unfolding avalanche thus breaking the
equivalence with a pure branching process. In other words, feedforward
loops may exist, meaning that a given node can be reached from a
unique seed by following different paths. This is particularly
relevant in undirected networks, where activity can attempt to go
backwards after any propagation event, following a reversible
link. This type of interference reduces the effective number of
independent offspring, breaking the BP analogy.

Second, networks in simulations are finite, implying that a finite
maximum degree exists, therefore the out-degree variance takes a
finite value; this implies that there should be crossovers to the
standard exponents for sufficiently large avalanche sizes and
durations.

Last but not least, 
some types of dynamics, even if taking place on
top of scale-free networks, do not really involve all neighbors of a
single node --as for example in the CP and IP processes-- and, thus,
have an effective offspring distribution narrowly distributed,
implying the emergence of standard MF exponents.

Numerous real-world systems have been investigated in terms of
avalanche statistics.  Prototypical examples include natural systems,
such as neuronal networks~\cite{beggs2003neuronal}, $\gamma$-ray
bursts~\cite{wang2013self} and earthquakes~\cite{bak2002unified}, as
well as socio-technical systems, such as power
networks~\cite{kinney2005modeling} and online social
media~\cite{nishi2016reply, qiu2017limited, wegrzycki2017cascade,
  lerman2010information}. Among them, some systems display critical
avalanche statistics consistent with the MF universality
class~\cite{beggs2003neuronal}.  However, there are many other systems
showing avalanche statistical properties that are not consistent with
the MF universality class. Examples can be found especially in the
literature studying information avalanches in online social media
where measured exponents for the power-law distribution of avalanche
size range from $\tau \simeq 4$~\cite{nishi2016reply}, to
$\tau \simeq 2.3$~\cite{wegrzycki2017cascade} and
$\tau \simeq 2$~\cite{qiu2017limited}.

Our analytical and numerical evidence supports the existence of an
extremely robust universality class at the interface between the absorbing
and the active phases of many popular models of avalanche dynamics.  Such
a universality class can be broken only at the expense of making
specific assumptions on the shape of the network underlying the
spreading model. We believe that it is imperative to understand why
there exist real systems that do not conform to such a class, and what
alternative hypotheses need to be made to account for their behavior.
In other words, a complete analytical theory --extending the
  approach presented here-- and accounting for all types of networks
  still needs to be constructed.

  As a final note, let us stress that our results reveal that
  constructing a dynamical model characterized by MF avalanche
  exponents is not a hard task.  Consequently, having a model that
  generates avalanche distributions with MF exponents --being these in
  agreement with some experimental observation-- does not constitute a
  sufficient evidence that the model is actually a sound one. Other
  dynamical aspects should be also used to validate the model.

  \acknowledgements{
F.R. and D.N. acknowledge support from the National Science Foundation
(CMMI-1552487). F.R. acknowledges support from the US Army Research
Office (W911NF-16-1-0104). M.A.M. acknowledges
the Spanish Ministry and Agencia Estatal de
  investigaci{\'o}n (AEI) through grant FIS2017-84256-P (European
  Regional Development Fund), as well as the Consejer{\'\i}a de
  Conocimiento, Investigaci{\'o}n y Universidad, Junta de
  Andaluc{\'\i}a and European Regional Development Fund (ERDF),
  ref. $A-FQM-175-UGR18$ and $SOMM17/6105/UGR$ for financial support.
    }


%

\end{document}